\documentclass[prd,10pt,twocolumn,tightenlines,superscriptaddress,numerical,notitlepage,showpacs,amsmath,amssymb,amsfonts,aps,longbibliography,floatfix
]{revtex4-1}
\usepackage{graphicx}
\usepackage[colorlinks=True,linkcolor=red,citecolor=blue,urlcolor=blue]{hyperref}
\usepackage{bm}
\usepackage{bbm}
\usepackage{array}
\usepackage{braket}
\usepackage{cancel}
\usepackage{xcolor}
\usepackage{slashed}
\usepackage{comment}
\usepackage{combelow}
\usepackage{chngcntr}
\usepackage{nicefrac}
\usepackage{bookmark}
\usepackage[T1]{fontenc}
\usepackage[english]{babel}
\usepackage[normalem]{ulem}

\newcommand{\beqn}{\begin{eqnarray}}
\newcommand{\eeqn}{\end{eqnarray}}
\newcommand{\beqs}{\begin{subequations}}
\newcommand{\eeqs}{\end{subequations}\\[-2mm]\noindent}

\newcommand{\bs}{\boldsymbol}

\newcommand{\tr}{\mathop{\mathrm{tr}}}

\newcommand{\GeV}{\,{\mathrm{GeV}}}

\definecolor{purple}{rgb}{0.8,0,0.6}
\definecolor{orange}{rgb}{1,0.55,0}
\definecolor{limegreen}{rgb}{0.2,0.8,0.2}

\allowdisplaybreaks

\begin{document}
\date{\today}

\author{M. N. Chernodub}
\affiliation{Institut Denis Poisson, CNRS UMR 7013, Universit\'e de Tours, Universit\'e d'Orl\'eans, Parc de Grandmont, Tours, 37200, France}
\affiliation{Department of Physics, West University of Timi\cb{s}oara, Bd. Vasile P\^arvan 4, Timi\cb{s}oara 300223, Romania}
\author{V. A. Goy}
\affiliation{Pacific Quantum Center, Far Eastern Federal University, 690922 Vladivostok, Russia}
\affiliation{Institute of Automation and Control Processes, Far Eastern Branch, Russian Academy of Science, 5 Radio Str., Vladivostok 690041, Russia}
\author{A. V. Molochkov}
\affiliation{Institut Denis Poisson, CNRS UMR 7013, Universit\'e de Tours, Universit\'e d'Orl\'eans, Parc de Grandmont, Tours, 37200, France}
\affiliation{Pacific Quantum Center, Far Eastern Federal University, 690922 Vladivostok, Russia}
\affiliation{Beijing Institute of Mathematical Sciences and Applications, Tsinghua University, 101408, Beijing, China}

\title{Acoustic phonons in a magnetized vacuum? First-principle lattice results \\ 
on the mass spectrum of the electroweak model in a strong magnetic field}

\begin{abstract}
We use numerical Monte Carlo simulations to determine the mass spectrum of the bosonic sector of the electroweak model in an external magnetic field of the electroweak-scale strength ($10^{20}\,{\rm T}$) at zero temperature. It is known that as the magnetic field gets stronger, the electroweak vacuum undergoes two consecutive crossover-type transitions, passing from (i) the conventional symmetry-broken homogeneous phase to (ii) an intermediate inhomogeneous vortex phase characterized by a (superconducting) condensate of electrically charged $W$ bosons and then to (iii) a homogeneous phase with a restored electroweak symmetry. We show that the spin component of the $W$ boson aligned with the direction of the magnetic field is the lightest excitation in all three phases. Its mass continuously decreases in the low-field broken phase and becomes very small in the intermediate phase. We argue that this nearly massless excitation corresponds to a Goldstone acoustic phonon mode associated with vibrations of the lattice of electroweak vortices. In the high-field symmetry-restored phase, where the vortices disappear, the lightest $W$ mass rises again. Neither Higgs nor $Z$ boson masses vanish across all studied phases and crossover transitions. 
\end{abstract}

\maketitle

\section{Introduction}

Within the Standard Model of particle physics, the electroweak sector provides a unified description of electromagnetic and weak interactions~\cite{Paschos2007}. The electroweak part of the model is a $SU(2)_L \times U(1)_Y$ gauge theory that couples leptons and quarks to a complex scalar doublet, the Higgs field, via the gauge vector fields~\cite{Langacker:2009my}. The present-day Universe resides in the low-temperature electroweak phase with a nonvanishing vacuum expectation value of the Higgs field that spontaneously breaks the electroweak symmetry down to its electromagnetic subgroup, $U(1)_{\rm e.m.}$~\cite{Pich:2005mk}. 

It is well known that thermal fluctuations tend to restore spontaneously broken symmetries~\cite{Kirzhnits1972, Weinberg1974}. As a result, the electroweak sector experiences a smooth crossover transition to the symmetry-restored phase~\cite{DOnofrio:2015gop} at a temperature of the order of the Higgs mass~\footnote{We use the natural units $\hbar = c = k_B = 1$ in the paper.}. It is believed that the restored phase appeared at least once in the course of the thermal evolution of our Universe~\cite{Mukhanov2005}. 

However, it is less known that a much richer phase structure of the electroweak vacuum occurs in the presence of a background magnetic (or hypermagnetic) field even at zero temperature. The magnetic fields of the relevant electroweak ($10^{20} \, {\rm T}$) scale might have been created at the cosmological electroweak phase transition in the first moments of the early Universe~\cite{Vachaspati:1991nm, Grasso:2000wj}. Such enormous fields were also suggested to exist even in the modern Universe in the vicinity of the magnetized black holes~\cite{Maldacena:2020skw, Ghosh:2020tdu, Gervalle:2024yxj}.

The nontrivial phase structure of the electroweak model can already be seen at the level of classical equations of motion. A constant Abelian magnetic field ${\boldsymbol{B}}$ coupled to a non-Abelian gauge sector of the electroweak model makes the lowest Landau level of charged vector bosons tachyonic once the field exceeds a critical strength given by the $W$ mass~$m_W$~\cite{Skalozub:1986gw, Ambjorn:1988fx, Ambjorn:1990xh, Ambjorn:1989hp, Ambjorn:1990xh, Ambjorn:1990ji, Ambjorn:1992ca}:
\begin{align}
	B^{\rm cl}_{c1} = m_W^2/e \qquad\ \mbox{[classical theory]}\,.
    \label{eq_Bc1_cl}
\end{align}
The tachyonic instability brings the homogeneous low-field vacuum with the Higgs condensate to the new phase characterized by the additional $W$ condensation and formation of an inhomogeneous hexagonal vortex-like state. The electroweak vacuum in this intermediate-field phase behaves as an anisotropic superconductor and, possibly, superfluid that can support dissipationless electric and neutral currents flowing along the magnetic field~\cite{Chernodub:2010qx, Chernodub:2011gs, Chernodub:2012fi}. 

At even stronger magnetic fields of the order of the Higgs mass $m_H$, 
\begin{align}
	B^{\rm cl}_{c2} = m_H^2/e \qquad\ \mbox{[classical theory]}\,,
    \label{eq_Bc2_cl}
\end{align}
the Higgs and $W$ condensates were suggested to be destroyed, and the classical analysis predicted that the gauge symmetry should be restored~\cite{Skalozub:1978, Ambjorn:1990xh, Ambjorn:1990ji, Ambjorn:1992ca, Chernodub:2012fi}. 

The proposed three-phase picture in the strong external magnetic field has been confirmed in the first-principles lattice simulations of the bosonic electroweak sector~\cite{Chernodub:2022gdo}. Numerically, the transitions between the phases appear to be smooth crossovers --- as contracted to genuine singular phase transitions --- with the following pseudocritical magnetic fields:
\begin{subequations}
\begin{align}
B_{c1} = &\, 0.68(5) m_W^2/e \,\simeq 0.8 \times 10^{20}\,{\rm T}\,,
\label{eq_Bc1_lat} \\ 
B_{c2} = &\, 0.99(2) m_H^2/e \; \simeq 2.7 \times 10^{20}\,{\rm T}\,,
\label{eq_Bc2_lat}
\end{align}
\label{eq_Bcs_lat}
\end{subequations}
\hskip -2mm where $m_W\simeq 80.4\GeV$ and $m_H\simeq 125.2\GeV$ are the masses of the $W$ and Higgs bosons, respectively~\cite{PhysRevD.110.030001}, and $e > 0$ is the elementary electric charge. One can readily observe that the first crossover emerges at a noticeably lower magnetic field~\eqref{eq_Bc1_lat} compared to the classical prediction~\eqref{eq_Bc1_cl} while the second crossover~\eqref{eq_Bc2_lat} perfectly matches the classical result~\eqref{eq_Bc2_cl}. 

Our work aims to establish, using the first-principle lattice simulations of the bosonic sector of the electroweak model, the low-lying part of the mass spectrum in all three zero-temperature phases across both crossover transitions~\eqref{eq_Bcs_lat} in the background magnetic field. 

The numerical simulations show that in the superconducting intermediate phase, the hexagonal order of the vortex lattice --- suggested by a solution to the classical equations of motion --- is not realized~\cite{Chernodub:2022gdo}. Instead, the vortices form a disordered vortex solid, which corresponds to strong vibrations of the vortices around their equilibrium hexagonal positions in the plane transverse to the direction of the magnetic field. This qualitative result points out a possible existence of the vibrational (phonon) excitations associated with the collective displacements of the vortex crystal in the electroweak vacuum. These phonon modes are similar to the acoustic and optical phonons associated with the vibrations of the vortex lines that appear in a mixed phase of type-II superconductors~\cite{DeGennes1964} and superfluids \cite{Tkachenko1966} (for a recent review, see Ref.~\cite{Rosenstein2010}).

In the electroweak model, the vibrating vortex lattice melts into a vortex liquid at the boundaries of the intermediate phase closer to both lower, $B = B_{c1}$, and higher, $B = B_{c2}$, pseudocritical values of the magnetic field. The vortex matter ceases to exist in the adjacent phases~\cite{Chernodub:2022gdo}. One of the aims of our work is to reveal a possible imprint of the phonon-like vortex excitation on the particle spectrum in the magnetized electroweak vacuum.

The structure of our paper is as follows. In Section~\ref{sec_electroweak_model}, we discuss the bosonic sector of the electroweak model in the continuum limit. Then, we briefly describe the discretized formulation of the model, referring an interested reader to Ref.~\cite{Chernodub:2022gdo} for further details. We introduce correlators that are used to determine the mass spectrum of the model and we also provide details on the parameters of the numerical simulations. Section~\ref{eq_analytical_vs_numerical} is devoted to a description of our numerical results confronted with the available analytical estimations for the evolution of the mass spectrum of Higgs, $W$ and $Z$ boson modes in the background magnetic field. In Section~\ref{sec_nature}, we discuss the nature of an almost-massless excitation found in the $W$-boson channel by contrasting it to photon-, magnon- and phonon-like quasiparticles in condensed matter systems. The last section is devoted to our conclusions.

\section{Electroweak model}
\label{sec_electroweak_model}

\subsection{Model in continuum limit}

We consider the bosonic part of the electroweak model with the Lagrangian,
\begin{align}
{\mathcal L}_{\rm EW}= &\, -\frac{1}{2} \tr (W_{\mu \nu} W^{\mu \nu}) - \frac{1}{4} Y_{\mu \nu} Y^{\mu \nu} 
    \nonumber \\
& + (D_\mu \phi)^\dagger (D^\mu \phi) - \lambda \bigl(\phi^\dagger \phi - v^2/2\bigr)^2,
\label{eq_LEW}
\end{align}
where $W_{\mu \nu}^a = \partial_\mu W_\nu^a - \partial_\nu W_\mu^a + i g \varepsilon^{abc} W_\mu^b W_\nu^c$ is the field strength of the SU(2) gauge field $W_\mu^a$ and $Y_{\mu \nu} = \partial_\mu Y_\nu - \partial_\nu Y_\mu$ is the field strength of the $U(1)_Y$ hypercharge gauge field $Y_\mu$. These vector fields interact with the complex scalar Higgs doublet $\phi \equiv (\phi_1,\phi_2)^T$ via the covariant derivative, 
\begin{align}
    D_\mu = \partial_\mu + \frac{i}{2} g W_\mu^a \sigma^a + \frac{i}{2} g' Y_\mu\,,
    \label{eq_D_mu}
\end{align}
where $\sigma^a$ ($a=1,2,3$) are the Pauli matrices. The ratio of the $U(1)$ and $SU(2)$ gauge couplings, $g'/g = \tan \theta_W$, defines the electroweak mixing (Weinberg) angle, $\theta_W$ which is fixed experimentally~\cite{CODATA2018} as $\sin^2 \theta_W \equiv 1 - m_W^2/m_Z^2 = 0.22290(30)$\,. The last term in Lagrangian~\eqref{eq_LEW} corresponds to the potential of the Higgs field doublet $\phi$, where $\lambda$ is the dimensionless self-coupling of the Higgs field. The only dimensionful parameter $v$ in the electroweak Lagrangian~\eqref{eq_LEW} sets the bare vacuum expectation value of the Higgs field, $\langle \phi \rangle = v$.

We consider the electroweak vacuum in the background of the hypermagnetic field ${\bs B}_Y = {\bs \nabla} \times {\bs Y}$ corresponding to the hypergauge field $Y^\mu = (Y^0, {\bs Y})$. In the broken phase, the hypermagnetic and magnetic fields are related:
\begin{align}
g' {\bs B}_Y = e {\bs B} \qquad {\mbox{[broken phase with $\langle \phi \rangle \neq 0$]}}\,,
\label{eq_BYB}
\end{align}
as it follows from the definition of the elementary electric charge, $e = g \sin\theta_W = g' \cos\theta_W = g g'/\sqrt{g^2 + g^{\prime 2}}$. In a symmetry-restored phase, the hypermagnetic field ${\bs B}_Y$ plays the role of a genuine field because the magnetic field $\bs B$ cannot be defined as the Higgs condensate vanishes and the direction of the electromagnetic symmetry group remains undetermined.

\subsection{Lattice setup and observables}

We used the Monte Carlo technique to simulate the electroweak theory in the lattice regularization~\cite{Chernodub:2022gdo.SM, Maas:2017wzi}. To generate an ensemble of field configurations, we employed the Hybrid Monte Carlo algorithm, which treats the gauge and matter fields simultaneously~\cite{Gattringer:2010zz}. The main results in our study were obtained on a lattice with the size $64\times 48^3$ ($N_t \times N_s^3$), where $N_t$ ($N_s$) is the size of the lattice in the temporal (spatial) direction(s). Additionally, the region around the first crossover transition was studied more precisely on a $72^4$ lattice to rule out the possible appearance of finite-volume effects. As we will see below, the volume effects are, indeed, rather small apart from a very narrow region around the first transition point. 

The bare lattice parameters were chosen to achieve the physical values of the masses and couplings of the electroweak sector at vanishing background field. The lattice spacing $a$ was fixed via the mass of the Higgs boson, $a m_H = 0.3815(6)$. Due to limitations in computer power, we have performed our calculations at the mentioned single value of the lattice spacing, which does not allow us to study the scaling of our results with the variation of the ultraviolet cutoff $\sim 1/a$. Nevertheless, given the good agreement of the numerically obtained phase diagram with the analytical expectations already noticed in Ref.~\cite{Chernodub:2022gdo} and, as we will see below, the reasonable match between analytical predictions and numerical results for the mass spectra in appropriate validity regions, we believe that our simulations run sufficiently close to the continuum limit.

To calculate the masses of the particle excitations in the background magnetic field, we followed the procedure described in detail in Ref.~\cite{Chernodub:2022gdo.SM}. We used the exponential fits of the appropriate two-point imaginary-time correlation functions 
\begin{align}
	  C_{(\mu)}^{\cal O}(\Delta \tau) = \sum_{\tau = 0}^{N_t - 1} \langle {\cal O}_{(\mu)}(\tau)  {\cal O}_{(\mu)}(\tau + \Delta \tau)\rangle \,, 
    \label{eq_C_l}
\end{align}
of the time-slice averaged operators, 
\begin{align}
	{\cal O}_{(\mu)}(\tau) = (1/V_s) \sum_{{\boldsymbol{x}} \in V_s} {\cal O}_{\boldsymbol{x},\tau;(\mu)}\,, 
    \label{eq_O}
\end{align}
where the sum of the appropriate point-like operator ${\cal O}_{\boldsymbol{x},\tau;(\mu)}$ is taken over the whole spatial volume $V_s$ at the fixed imaginary time $\tau$. The index ``$\mu$'' is reserved for vector quantities that will be discussed below.

The two-point functions~\eqref{eq_C_l} were fitted by the hyperbolic function of the imaginary time distance $\Delta \tau$:
\begin{align}
  C_{\rm fit}^{\cal O}(\Delta \tau) = c_0 \cosh \bigl[m_{\cal O} \bigl(\Delta \tau - N_t/2\bigr) \bigr]\,,
   \label{eq_C_l_fit}
\end{align}
where $c_0$ and $m_{\cal O}$ are the fitting parameters. Function~\eqref{eq_C_l_fit} respects the periodic boundary conditions, $C(\Delta \tau) = C(N_t - \Delta \tau)$, along the imaginary time $\tau$ direction. The mass parameter $m_{\cal O}$ corresponds to the (double) mass of the lightest excitation that has an overlap with the state created by the operator ${\cal O}$.

For the Higgs mass, we choose the squared modulus of the Higgs field, ${\cal O}_{\boldsymbol{x}, \tau} = |\phi_{{\boldsymbol{x},\tau}}|^2$ in the time correlator~\eqref{eq_C_l} of the bulk operators \eqref{eq_O}. Then the fitting function~\eqref{eq_C_l_fit} gives us the mass of the Higgs field $m_{\cal O} = M_H$ in the broken and superconducting/vortex phases with the Higgs condensate. In the third, symmetry-restored phase, the Higgs condensate is absent, and, therefore, the quantity $m_{\cal O}$ corresponds to a double mass of the scalar particle in this phase.

The masses of the $W$ and $Z$ bosons are calculated using, respectively, the operators ${\cal O}_{\boldsymbol{x}, \tau; \mu}  = W_\mu, Z_\mu$, where the discretized versions of these fields can be found in Ref.~\cite{Chernodub:2022gdo.SM}. In Eq.~\eqref{eq_C_l}, we summed over the correlators with the transverse Lorentz indices, $\mu = 1,2$ and we considered separately the longitudinal field with $\mu = 3$. To reduce ultraviolet noise for the $W$- and $Z$-boson masses, we applied the spatial APE-smearing procedure~\cite{Durr:2004xu} with parameters $\alpha_{\mathrm{APE}}=0.5$ and $n_{\mathrm{APE}}=100$ for gauge fields. After smearing, we fixed a maximal tree gauge for the $U(1)$ gauge field and unitary gauge for the $SU(2)$ gauge field, then calculated the two-point function. 

On a $64\times 48^3$ and $72^4$ lattices, we used $(1.7{-}3.0)\times 10^6$ and $(0.5{-}1.0)\times 10^6$ configurations, respectively. The error estimation of the observables is performed after taking into account the autocorrelations in the Markov chains. Below, we discuss the theoretical expectations and the numerical results for each studied field. 

\begin{figure}
  \centering
  \includegraphics[width=1.0\linewidth]{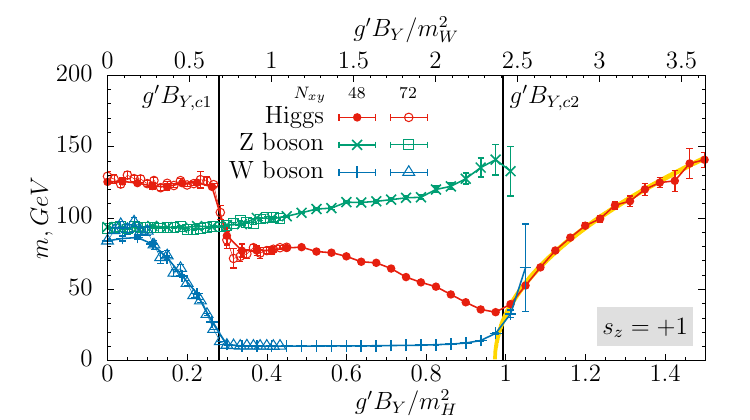}\\[3mm]
  \includegraphics[width=1.0\linewidth]{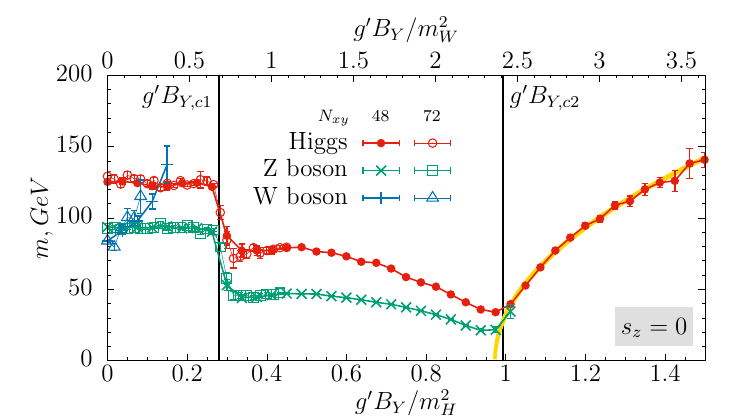}
  \caption{Masses of $Z$ and $W$ bosons with $s_z = +1$ (top) and $s_z = 0$ (bottom) projections of the spin on the magnetic field axis as a function of magnetic field strength given in terms of the Higgs (the lower axis) and $W$ (the upper axis) masses for different lattice sizes. The mass of the Higgs field is shown on both plots for reference. Two vertical solid lines mark the pseudocritical transitions~\eqref{eq_Bcs_lat} obtained in Ref.~\cite{PhysRevD.110.030001}. The yellow solid line in each plot represents the best fit of the Higgs mass by function~\eqref{eq_Higgs_high_fit}.}
  \label{fig:M}
\end{figure}

\section{Analytical vs. numerical results}
\label{eq_analytical_vs_numerical}

\subsection{Free vector bosons in a magnetic field}

The first transition at the critical field~\eqref{eq_Bc1_lat} is associated with the tachyonic instability and the condensation of the charged $W$ bosons. The existence of this transition can already be established at the tree (classical) level~\cite{Skalozub:1986gw, Ambjorn:1988fx, Ambjorn:1990xh, Ambjorn:1989hp, Ambjorn:1990xh, Ambjorn:1990ji, Ambjorn:1992ca} based on the following straightforward arguments. 

The Landau energy levels of a free vector (spin $S = 1$) particle with the mass $m$, gyromagnetic ratio $g_m$ and electric charge $e$ in a uniform static external magnetic field $\boldsymbol{B} = B\,\hat{\boldsymbol{z}}$ are labeled by a non-negative integer $n = 0,1, \dots$ and the spin projection $s_z = - 1, 0, 1$ on the direction of the magnetic field:
\begin{align}
  E^2_{n, s_z}(k_z) = m^2 + (1 + 2 n - g_m s_z) eB + k_z^2\,,
  \label{eq_En}
\end{align}
where $k_z$ is the momentum of the particle in the direction of the magnetic field (for simplicity of notations, we used the convention $eB > 0$). Notice that the same result~\eqref{eq_En} is also valid for Dirac fermions (with $s_z = \pm 1/2$) and scalar particles (with $s_z = 0$). Similarly to the Dirac fermions, the $W$ bosons in the electroweak theory have the gyromagnetic ratio $g_m = 2$. In the case of the $W$ bosons, the large gyromagnetic ratio --- as opposed to the trivial $g_m = 1$ for a Proca field --- appears due to the non-Abelian origin of these vector particles. 

Let us now focus on a single-charged state of the vector boson, say $W_\mu \equiv W_\mu^-$. A massive vector field possesses three spin polarizations, $s_z = -1, 0 , +1$. In the presence of the background magnetic field, it is convenient to use a basis of eigenstates of the spin projection on the $z$ axis~$S_z$:
\begin{align}
  &{\bf e}^{(\pm)} = \frac{1}{\sqrt{2}}(1, \pm i, 0)\,, \qquad {\bf e}^{(0)} = (0,0,1)\,,
\end{align}
which satisfy $S_z\,{\bf e}^{(s_z)} = s_z\,{\bf e}^{(s_z)}$ with $s_z = \pm 1, 0$. The spatial part of the $W$-boson field can be decomposed as
\begin{align}
 {\boldsymbol{W}} = W^{(+)}\,{\bf e}^{(+)} + W^{(0)}\,{\bf e}^{(0)} + W^{(-)}\,{\bf e}^{(-)}\,.
\end{align}
The modes $W^{(\pm)}$ correspond to circularly polarized transverse excitations in the $(x,y)$ plane with spin projections $s_z = \pm 1$, implying that $W^{(\pm)}_{x}$ and $W^{(\pm)}_y$ components are generally nonzero while the longitudinal component always vanishes, $W_z = 0$. For the spin projection $s_z=0$, the orientation of the field components is normal to the $s_z = \pm 1$ projections since $W^{(0)}_{x} = W^{(0)}_{y} = 0$, while the $W^{(0)}_{z}$ component parallel to the field is nonzero. These polarization branches have the following energies: 
\begin{subequations}
\begin{align}
 E_{n,+}^2(k_z) & = m_W^2 + (2n-1)\,eB + k_z^2 \,, \label{eq_E_plus} \\
 E_{n,0}^2(k_z) & = m_W^2 + (2n+1)\,eB + k_z^2 \,, \label{eq_E_zero} \\
 E_{n,-}^2(k_z) & = m_W^2 + (2n+3)\,eB + k_z^2 \,. \label{eq_E_minus}
\end{align}
\end{subequations}

For the low Landau levels ($n=0, 1, \dots$) at a zero longitudinal momentum ($k_z=0$), one finds the following hierarchy of the masses (with $eB > 0$):
\begin{subequations}
\begin{align}
 E_{0,+}^2(0) & = m_W^2 - eB\,, \label{eq_M_plus}\\
 E_{0,0}^2(0) = E_{1,+}^2(0) & = m_W^2 + eB\,, \label{eq_M_zero} \\
 E_{0,-}^2(0) = E_{1,0}^2(0) = E_{2,+}^2(0) & = m_W^2 + 3eB \label{eq_M_minus}\,,
\end{align}
\label{eq_masses}
\end{subequations}
and so on. 

Thus, one of the transverse modes, with spin aligned along the field ($s_z=+1$), becomes gradually lighter as the magnetic field increases. Its mass~\eqref{eq_M_plus} vanishes at the first critical field~\eqref{eq_Bc1_cl} and turns tachyonic for $e B > m_W^2$, signaling the onset of the instability of the trivial vacuum state against the $W$-boson condensation~\cite{Skalozub:1986gw, Ambjorn:1988fx, Ambjorn:1990xh, Ambjorn:1989hp, Ambjorn:1990xh, Ambjorn:1990ji, Ambjorn:1992ca}.

In contrast, the longitudinal polarization ($s_z=0$) and the opposite transverse polarization ($s_z=-1$) acquire larger effective masses, Eqs.~\eqref{eq_M_zero} and \eqref{eq_M_minus}, respectively. This observation implies that, in the background magnetic field, one should distinguish the circularly polarized modes with the components in the $(x,y)$ plane and the longitudinally polarized modes along the nonvanishing component along the $z$ axis.

The masses of the $W$ and $Z$ vector bosons in the electroweak model depend not only on their polarizations with respect to the background magnetic field but also on the expectation value of the Higgs field, which, in the absence of the background magnetic field, gives masses to all particles in the Standard Model. Below, we continue our discussion by addressing first the mass of the Higgs boson and coming back to the properties of the vector bosons afterwards.

\subsection{Higgs field}

\subsubsection{Analytical expectations}

The mass of the Higgs field in the absence of the background magnetic field, $m_H = \sqrt{2 \lambda} v$, depends on the self-coupling $\lambda$ and the vacuum expectation value $v \equiv |\langle \phi \rangle|$. One can use this formula for a qualitative estimation of the behavior of the Higgs mass in the background hypermagnetic field. 

In both broken (lower-field) phases at $B < B_{c2}$, only the $U(1)_{\rm e.m.}$ magnetic component $B$ of the hypermagnetic field $B_Y$ survives in the vacuum. Since the Higgs particle is not electrically charged, the Higgs field does not interact with the magnetic field, and consequently, it does not couple directly to the hypermagnetic field via a covariant derivative in these two phases. Therefore, the only effect of the hypermagnetic field on the Higgs mass comes indirectly, from classical interactions with other fields and via the quantum loops. Due to the same reason, we expect that in all three phases, the self-coupling constant $\lambda$ is not affected by the strength of the background magnetic field at the classical level.

According to analytical considerations based on the classical equations of motion, the expectation value $\langle \phi \rangle$ stays constant in the low-field, broken phase ($B < B_{c1}$) until the field reaches the first critical field~\eqref{eq_Bc1_cl}. At the intermediate values of the field, superconducting phase ($B_{c1} < B < B_{c2}$), the $W$ condensate diminishes the expectation value of the Higgs field $\langle \phi \rangle$ because the $W$ condensate adds a positive contribution to the (quadratic) curvature of the Higgs potential~\cite{Ambjorn:1990xh, Ambjorn:1990ji, Ambjorn:1992ca, Chernodub:2012fi}. The Higgs expectation value $\langle \phi \rangle$ should then vanish at the second critical field~\eqref{eq_Bc2_cl}, where the quadratic curvature becomes zero.

In the high-field restored phase ($B > B_{c2}$), the hypermagnetic field is no more screened. The hypergauge field $Y_\mu$ couples directly to the Higgs field via the covariant coupling~\eqref{eq_D_mu}. Given that the Higgs particle carries the hyperelectric charge $g_Y = g'/2$, the behavior of the mass of the Higgs field can be learned from the properties of the free scalar field in the background of the (hyper)magnetic field, 
\begin{align}
  m^{\rm cl}_H(B_Y)=\sqrt{M_{H}^2 + g_Y B_Y} \qquad\ \mbox{[a free scalar]}\,,
  \label{eq_Higgs_free}
\end{align}
where we take $g_Y B_Y > 0$ for simplicity. Relation~\eqref{eq_Higgs_free} follows from Eq.~\eqref{eq_En} for a scalar (spin-zero, $s_z = 0$) particle at the lowest Landau level ($n=0$) at a zero longitudinal momentum ($k_z = 0$) with the substitutions $e B \to g_Y B_Y$ for the magnetic field and $m \to M_{H}$ for the zero-field ($B_Y = 0$) mass.

Summarizing, the classical theory predicts that the Higgs mass should be constant in the broken phase at $0 < B < B_{c1}$ and it should gradually diminish with the rising magnetic field in the superconducting vortex phase at $B_{c1} < B < B_{c2}$. Then, it should vanish at $ B = B_{c2}$ before rising again in the restored phase, $B > B_{c2}$, following the approximate profile of Eq.~\eqref{eq_Higgs_free}.

\subsubsection{Numerical analysis}

Our numerical data shown in Fig.~\ref{fig:M} partially aligns with the analytical expectations described above. The Higgs mass appears to be largely independent of the magnetic field in the broken phase. Then, it falls down sharply to a non-zero value at the first critical line. As the field strength increases further, the mass exhibits a plateau and then continues a smooth decline in the superconducting phase, reaching a minimum with a non-zero mass at the second critical field, and then rising again in the third phase.

Assuming a non-vanishing Higgs mass at the second phase transition ($B = B_{c2}$), one gets the following refined dependence~\eqref{eq_Higgs_free} appropriate for the third, symmetry-restored phase at $B > B_{c2}$:
\begin{align}
  m^{\rm (fit)}_H(B_Y)=\sqrt{m^2_{H^*} + G\, (g' B_Y - m_H^2)} 
  \,.
  \label{eq_Higgs_high_fit}
\end{align}
Here $m_{H^*} = m_H(B = B_{c2})$ is the mass of the Higgs field at the position of the second critical magnetic field~\eqref{eq_Bc2_cl}, which corresponds to $g' B_{Y,c2} = m_H^2$ via relation~\eqref{eq_BYB} that should be valid at (the lower-field side of) the second critical field, $B = B_{c2}$. The phenomenological prefactor $G$ in Eq.~\eqref{eq_Higgs_high_fit} is used to incorporate quantum fluctuation effects that can modify the strength of the interaction with the hypermagnetic field $B_Y$. We treat $m_{H^*}$ and $G$ as variable parameters to fit the data by Eq.~\eqref{eq_Higgs_high_fit}. 

It appears that the phenomenological function~\eqref{eq_Higgs_high_fit} perfectly matches the Higgs mass in the high-magnetic field phase at $B >B_{c2}$. The best-fit curve is shown by the solid yellow line in Fig.~\ref{fig:M}.

We find that at the second transition point~\eqref{eq_Bc2_lat}, the mass of the Higgs boson is well above zero, $m_{H^*} = 28(1)\GeV$. This result is in contrast with the theory, as one expects that the Higgs mass vanishes in the second-order phase transition~\cite{Ambjorn:1990xh, Ambjorn:1990ji, Ambjorn:1992ca, Chernodub:2012fi}. Our fit also shows that the electromagnetic coupling is getting substantially enhanced in the interacting theory, giving us $G = 2.44(4)$ instead of the expected $G = 1/2$ in the free scalar case~\eqref{eq_Higgs_free}.


\subsection{$\bf{W}$ boson}

The expected analytical properties of the $W$ boson in the low-field phase with $B < B_{c1}$ have already been discussed above. Here, we summarize our numerical data for the masses of the $W$ boson states in all three phases and then discuss a theoretical interpretation of these first-principle results.

\subsubsection{$s_z = + 1$ spin state of $W$ boson}

In the numerical simulations, we studied the mass spectrum of the $W$ boson excitations separately for the transverse and longitudinal components of the $W$ field. Our numerical method picks up only the lightest excitation in each channel. Therefore, according to Eq.~\eqref{eq_masses}, the transverse correlation functions give us the $s_z = + 1$ state of the $W$ boson (the lightest excitation possessing the transverse components), while the longitudinal correlation function gives us the mass of the $s_z = 0$ component (the lightest excitation that has a nonvanishing longitudinal component).

The mass of the $s_z = + 1$ $W$-state is shown, as a function of the background magnetic field, by the blue line in the top plot of Fig.~\ref{fig:M}. In agreement with the analytical prediction of Ref.~\eqref{eq_M_plus}, in the first, low-field phase with $B < B_{c1}$, the mass decreases with the increase of the magnetic-field strength~$B$.

\begin{figure}
  \centering
  \includegraphics[width=1.0\linewidth]{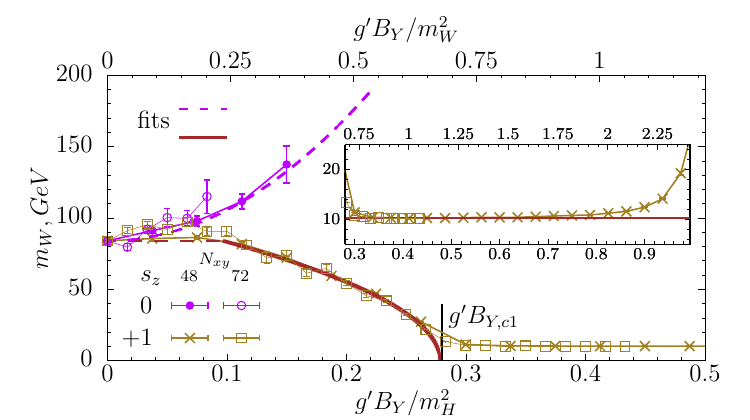}
  \caption{The masses of the $s_z = +1$ and $s_z = 0$ spin states of the $W$ boson field in the broken and superconducting phases. The theory-inspired fit~\eqref{eq_W12_fit} of the $s_z = +1$ mass is represented by the solid brown line, which hits zero at the point $B = B_{c1}$ of the first phase transition~\eqref{eq_Bc2_lat}. The dashed magenta line shows the parabolic fit~\eqref{eq_fit_mz0} of the mass of the $s_z = 0$ state. The plateau in the mass of the $s_z = + 1$ state in the intermediate, superconducting phase ($B_{c1} < B < B_{c2}$) appears at value~\eqref{eq_W_plateau}, shown in the inset by the horizontal solid line.}
  \label{fig:MW}
\end{figure}

A low-field corner of the phase diagram is zoomed in Fig.~\ref{fig:MW}. It appears that the behavior of the mass of the $s_z = +1$ state of the $W$ boson can be well described by the following fitting function:
\begin{align}
  m^{\rm (fit)}_{W_{s_z = +1}}(B) = \gamma \sqrt{\alpha m_W^2 - eB}\,,
  \label{eq_W12_fit}
\end{align}
with the best fit parameters:
\begin{align}
    \alpha = 0.676(4)
    \qquad 
	\gamma=1.57(2)\,.
  \label{eq_W12_fit_2}
\end{align}
While the general form of the function~\eqref{eq_W12_fit} follows the classical result~\eqref{eq_M_plus}, the fact that the values of $\alpha$ and $\gamma$ are different from unity implies that the quantum corrections are essential. In particular, Eq.~\eqref{eq_W12_fit_2} implies that the first phase transition should be realized at a critical value $B_{c1} = 0.676(4) m_W^2/e$ of the magnetic field, which corresponds to the root of expression~\eqref{eq_W12_fit}. The quoted value of the best-fit parameter $\alpha$ perfectly matches the first critical field~\eqref{eq_Bc1_lat} reported earlier in Ref.~\cite{Chernodub:2022gdo}. Also, since the transition happens earlier, the slope of the mass should be steeper, implying $\gamma > 1$ in an agreement with the data shown in Eq.~\eqref{eq_W12_fit_2}.

It is worth stressing that the fitting function~\eqref{eq_W12_fit} does not work in the very vicinity of the transition point. This observation is consistent with the crossover nature of the transition. Indeed, our data indicate that the lowest mass of the $W$ boson at the first transition point $B = B_{c1}$ does not vanish, being of the order of $m_W(B = B_{c1}) \simeq 12 \GeV$. This result appears to be in a qualitative agreement with the earlier theoretical considerations of Ref.~\cite{Skalozub2014}, which claims that quantum corrections make the mass positive. 

At this point, one should notice that our analysis has been made at a single value of the lattice spacing, and the finite-size analysis with respect to the variation of the ultraviolet cutoff has not been performed. However, given the robustness of the lattice results with the volume variation --- which also points out the crossover nature of the transition --- we believe that the lightest $W$-boson state will remain nonzero even in the continuum limit. 

At higher magnetic fields, the mass of the $s_z = + 1$ state of the $W$ field has two more features. Figures~\ref{fig:M} and \ref{fig:MW} show that in the second, moderate-field phase at $B_{c1} < B < B_{c2}$, the $W$ mass stays (i) very small, $m_W(B) \ll m_W \equiv m_W(B=0) \simeq 84 \GeV$ and (ii) almost independent of the field $B$. The plateau (``plt'') in the $W$ mass appears at the value
\begin{align}
    m_{W, {\rm plt}} = 10.2(2) \GeV\,, \qquad [B_{c1} \lesssim B \lesssim B_{c2}]\,,
    \label{eq_W_plateau}	
\end{align}
shown by the solid brown line in the inset of Fig.~\ref{fig:MW}. 

As the field strength approaches the third, high-field phase at $B > B_{c2}$, the $s_z = + 1$ component of the $W$-boson mass starts to grow up again with the increase of the magnetic strength. Due to strong field fluctuations, we were able to calculate the masses of this and other vector states only in the lowest-field regime of this phase, while at higher fields, the data has too large statistical errors to give a reliable result.

We discuss interpretations of these findings in Section~\ref{sec_nature} below.

\subsubsection{$s_z = 0$ spin state of $W$ boson}

According to Eq.~\eqref{eq_M_zero}, the longitudinal ($s_z = 0$) component of the $W$-boson field should grow linearly with the increasing magnetic field strength. Qualitatively, this feature is indeed observed in the bottom plot of Fig.~\ref{fig:M}. However, the quantitative behavior of the $s_z = 0$ state is inconsistent with the prediction given by Eq.~\eqref{eq_M_zero}. The best fit appears to be a quadratic function of the magnetic field
\begin{align}
    m^{\rm (fit)}_{s_z = 0}(B) = m_W \biggl[1 + \alpha \biggl(\frac{eB}{m_H}\biggr)^2 \biggr]\,,
    \label{eq_fit_mz0}
\end{align}
where $m_W = m_W(B=0)$ is the $W$ mass in the absence of the field (shown by the dashed brown line in Fig.~\ref{fig:MW}) and $\alpha = 26(5)$. The best fit is shown in Fig.~\ref{fig:MW} by the dashed magenta line.


\subsection{$\bf{Z}$ boson}

The mass of the $s_z = +1$ spin state of the $Z$ boson obtained with the help of the correlations of the transverse components of the $Z$ field is shown in the top plot of Fig.~\ref{fig:M}. The mass of its $s_z = 0$ counterpart, determined from the longitudinal correlators, is shown in the bottom plot of the same figure. 

At the tree level, the $Z$ boson does not couple to the background magnetic field. Therefore, its mass  should be given, up to quantum corrections, by the expectation value of the Higgs field, which is also related to the Higgs mass in the symmetry-broken phases ($B < B_{c2}$). One can observe that the $s_z = 0$ mass indeed qualitatively follows the behavior of the Higgs mass in these two phases, mimicking all its characteristic features at the transition points. 

The same arguments could also be applied to the $s_z = +1$ state. However, in an apparent contradiction with our theoretical expectations, the top plot of Fig.~\ref{fig:M} shows that the mass of the $s_z = +1$ spin state does not follow the Higgs mass. Moreover, it stays almost constant in the low-field phase ($B < B_{c1}$) and starts rising very slowly and monotonically in the intermediate phase ($B_{c1} < B < B_{c2}$). The reason for such behavior could possibly be rooted in the (weak) coupling of the $s_z = +1$ spin projection of the $Z$ boson field to the inhomogeneities of the $W$ condensate~\cite{Chernodub:2012fi} as the $W$ condensate vanishes in the low-field phase and rises in the intermediate-field phase. We discuss a consequence of the latter result at the end of the next section.

\section{Origin of the lightest $W$-boson mode in the superconducting phase}
\label{sec_nature}

Having established the spectrum of all accessible light excitations across various values of magnetic field, one readily notices the striking feature: in the superconducting phase at intermediate magnetic fields, $B_{c1} < B < B_{c2}$, the $s_z = +1$ state of the $W$ boson becomes an almost massless excitation with its mass being much smaller than a typical electroweak mass scale~\eqref{eq_W_plateau}. All other excitations are substantially more massive. 

The plateau mass~\eqref{eq_W_plateau} does not depend on the spatial volume of the system, indicating that its value is not limited by the longest wavelength in the system. However, the value of the plateau may be influenced by the ultraviolet artifacts that were not studied in this paper due to the limitations in the computing power. However, even with these limitations, one can definitely claim that the calculated mass of the $W$ boson state is very small at the scale of other electroweak masses. 

The soft (massless or nearly massless) modes could appear due to a spontaneous breaking of continuous symmetries associated, for example, with an internal symmetry group or with external symmetries such as translation or rotations of the spacetime. Having this observation in mind, let us ask, what is the nature of the nearly massless $W$ state observed in our simulations?

To proceed further, let us first recall some facts about the intermediate, superconducting phase $(B_{c1} < B < B_{c2})$. As the magnetic field exceeds the first critical magnetic field~\eqref{eq_Bc1_cl} or \eqref{eq_Bc1_lat}, the energy barrier of the $W^+W^-$ pair creation disappears, and the magnetic field facilitates the production of the $W^+$ and $W^-$ pairs by quantum fluctuations. The emerging pairs condense and form an electrically neutral and inhomogeneous condensate made of an equal number of positively and negatively charged $W$ bosons. This intermediate phase has several distinctive features at the level of classical equations of motion~\cite{Skalozub:1986gw, Ambjorn:1988fx, Ambjorn:1990xh, Ambjorn:1989hp, Ambjorn:1990xh, Ambjorn:1990ji, Ambjorn:1992ca}: 
\begin{itemize}
    \item[(i)] {\bf Electric charge.}  As in an ordinary superconductor, the condensate couples to the electromagnetic field through a minimal coupling, implemented via a covariant derivative. Consequently, although the condensate is electrically neutral, it can still exhibit a superconducting-like electromagnetic response. This response is strongly anisotropic, reflecting the underlying structure of the condensate and leading to markedly different electromagnetic properties along distinct spatial directions~\cite{Chernodub:2010qx}.

\item[(ii)] {\bf Spin and magnetic moment.} The background magnetic field couples to the magnetic dipole moments of the $W$ bosons, orienting them along the magnetic-field axis. The magnetic moments of $W^+$ and $W^-$ bosons are arranged oppositely with respect to their spins, implying that the total spin polarizations of the $W^+$ states and the $W^-$ states in the condensate are equal in magnitude and opposite in their directions. Thus, the net spin polarization of the condensate is zero. However, the $W$ condensate carries a nonzero magnetic moment density aligned with the external field, thus making it somewhat similar to a paramagnetic material~\cite{Ambjorn:1992ca}. 

\item[(iii)] {\bf Vortex lattice.} Geometrically, the condensate represents a spatially inhomogeneous structure that incorporates a lattice of infinitely long parallel vortices. The vortices form a hexagonal lattice in the normal plane perpendicular to the direction of the magnetic field. The numerical simulations in Ref.~\cite{Chernodub:2022gdo} have revealed that the quantum fluctuations destroy the hexagonal lattice order by forcing the vortices to move/vibrate with a large amplitude around their equilibrium positions (as demonstrated by the video in Ref.~\cite{Chernodub:2022gdo.SM}). 
\end{itemize}

With all these observations, let us make an attempt to determine the physical origin of the low-mass state~\eqref{eq_W_plateau} that appears in the $W$ channel.

First of all, the lightest excitation cannot be related to a Goldstone boson associated with the $U(1)_{\rm e.m.}$ symmetry. This Goldstone boson is represented by the scalar phase of the condensed transverse $W$ field, which would stay massless in the absence of the gauge field. In the electroweak model, however, this scalar particle is absorbed into the longitudinal component of the vector photon field, making the photon a massive particle in the superconducting phase via a (partial) Anderson-Higgs mechanism (the longitudinal $W$ components that do not enter the $W$ condensate remain massless)~\cite{Chernodub:2012fi}. Therefore, the lightest excitation coupled to the $W$ field cannot be associated with the electromagnetic $U(1)_{\rm e.m.}$ subgroup and it cannot be a photon. 

The lightest excitation cannot be associated with the spin of the condensed $W$ state. Indeed, as we argued above, the spin of the condensed state is vanishing, which excludes a spin wave as a potential candidate. 

On the other hand, the condensed $W$ state acquires a finite magnetization, indicating that the ground state supports a nontrivial magnetic order. This observation motivates a systematic investigation of its low-energy collective spectrum, in particular the possible emergence of magnon-like excitations corresponding to coherent fluctuations of the magnetization about the ordered background.
However, a paramagnet does not host true magnons because any paramagnet --- contrary to a ferromagnetic material~\cite{Kittel1949, Prabhakar2008} --- has no long-range order and no spontaneously broken spin-rotation symmetry. Moreover, if the magnons existed, the external magnetic field would fully polarize the system, and the breaking of the rotation symmetry, should it be the case, would lose its spontaneous nature. Therefore, the symmetry considerations exclude the presence of the magnon-like excitations in the spectrum of the system.

Since the charge, spin and magnetization characteristics of the condensed $W$ state cannot lead to a massless (or small-mass) excitation, the only possibility that remains should be associated with the dynamics of the vortices themselves. The corresponding propagating modes do indeed exist in the crystalline condensed matter environments. For example, a vortex lattice in a type-II superconductor is known to host the acoustic modes associated with transverse and longitudinal vibrations of the vortices about their equilibrium positions~\cite{DeGennes1964, Tkachenko1966, Rosenstein2010}. Since the vortices are embedded into the $W$ condensate, the vortex oscillations should mediate the interaction between the $W$ fields in a form of a hybridization of the quantum fluctuations of the $W$ field and those induced by the vortex vibrations. 

In other words, the correlators of the $W$ fields should get a contribution from the phonon-exchange channel. The displacements of vortices around the hexagonal equilibrium positions observed in numerical simulations of Ref.~\cite{Chernodub:2022gdo} could, therefore, be associated with these acoustic modes. 

The existence of the acoustic modes in the vortex lattices is a common phenomenon in type-II superconductors~\cite{Rosenstein2010}. These modes become gapped and dissipative as the vortex lattice melts~\cite{Brandt1995}. To put this statement into the perspective of our work, we recall that the vortex lattice in the magnetized electroweak vacuum melts close to the transition points within the intermediate phase, at $B \to B_{c1}^+$ and $B \to B_{c2}^-$~\cite{Chernodub:2022gdo}. In accordance with the observed vortex melting, our data in Fig.~\ref{fig:MW} shows that the lightest $W$ excitation becomes more massive as we approach any of these boundary points, thus supporting our interpretation of the lightest $W$ mode.

Thus, we conclude that the magnetized electroweak vacuum literally conducts sound modes associated with the elastic vibrations of the vortex lattice, in close similarity with acoustic modes in ordinary crystalline materials and superconducting/superfluid vortex lattices.  

Before closing this section, it is worth mentioning another outcome of our simulations. An analysis of the classical ground state of the theory indicates that the interaction of the condensed $W$ field and the $Z$ bosons leads to an instability of the neutral $Z$ bosons against their condensation~\cite{Chernodub:2012fi}. Moreover, it appears that both the charged $W$ condensate and the neutral $Z$ condensate support, respectively, charged and neutral dissipationless currents along the direction of the magnetic field. The currents are governed by an anisotropic London-like equations~\cite{Chernodub:2012fi, Chernodub:2012bj}. For the neutral condensate, these observations imply the emergence of superfluidity and, in turn, the inevitable presence of the phonon-like modes, the long-wavelength sound excitations of the superfluid. 

Should the massless superfluid phonons exist, they would not be left unnoticed in our simulations since the mass of the $s_z = +1$ state of the $Z$ boson field would be vanishing in the presence of a massless phonon mode. As it is not the case, we conclude that the intermediate phase does not possess a massless phonon associated with superfluidity, and consequently, the superfluidity may be absent in the strongly magnetized vacuum of the electroweak model. One could suggest that the superfluid state of the $Z$ bosons, which is supported by the classical analysis, is destroyed by strong vibrations of the vortices in the superconducting $W$ condensate.

\section{Conclusions}

Using first-principle numerical simulations, we established the mass spectrum of the bosonic sector of the electroweak theory in the background of the strong (hyper)magnetic field at zero temperature. In agreement with the earlier theoretical analysis~\cite{Nielsen:1978rm, Skalozub:1978, Skalozub:1986gw, Ambjorn:1988fx, Ambjorn:1989hp, Ambjorn:1988gb, Ambjorn:1990xh, MacDowell:1991fw, Salam:1974xe, Linde:1975gx}, supported by our previously reported numerical simulations~\cite{Chernodub:2022gdo}, the evolution of the mass spectrum with the increase of magnetic field confirms the existence of two crossover-type transitions~\eqref{eq_Bcs_lat} between three distinct phases. 

The first phase is the standard broken electroweak phase. The second phase is characterized by the inhomogeneous $W$-boson condensate with embedded vortices parallel to the background magnetic field. In the first two phases, the electroweak symmetry is broken, and the hypermagnetic field is related to the usual magnetic field. In the third, symmetry-restored phase, the magnetic field cannot be defined and we use the hypermagnetic field as a control parameter. 

The mass spectrum of the bosonic fields shows a remarkable evolution as the strength of the field increases (Fig.~\ref{fig:M}). Neither Higgs nor $Z$ boson masses vanish across all studied phases and crossover transitions. The mass of the Higgs field stays largely constant in the broken phase. It sharply drops at the first crossover transition by almost a half, then slowly diminishes again in the mixed (superconducting/vortex) phase, reaching about one-third of the zero-field value at the second phase transition before starting to rise rapidly in the restored phase. 

The presence of the background magnetic field leads to an anisotropy in the masses of the vector fields that acquire the strong dependence on the value of the spin projection $s_z$ onto the magnetic field axis $\boldsymbol{B}$. The $s_z = 0$ component of the $Z$ boson follows qualitatively the complicated evolution of the Higgs mass, while the $s_z = + 1$ component grows very slowly across the first and the second phases.

The mass of the $s_z = + 1$ state of the $W$-boson field drops in the broken phase, stays (nearly) zero in the superconducting phase (Fig.~\ref{fig:MW}), and then grows again in the restored phase. The mass of the $s_z = 0$ component of the $W$ boson increases so rapidly that we can only confirm its value in the first, symmetry-broken phase. Presumably for the same reason, masses of all other vector fields cannot be determined deeper in the third, symmetry-restored phase. 

We interpret the nearly vanishing mass of the $s_z = +1$ state of the $W$ boson as evidence of the existence of a phonon-like vector excitation associated with the mechanical vibrations of the vortex lattice in the intermediate, superconducting phase of the vacuum. We have excluded the photon-like Goldstone modes from the candidate list because those modes become massive via the Anderson-Higgs mechanism. Also, the massless magnon-like modes associated with the magnetization of vacuum cannot appear in the spectrum because the direction of the magnetic polarization in the system emerges as a result of the explicit, as opposed to spontaneous, rotational symmetry breaking. 

The phonon modes in the lattice of electroweak vortices are expected to be similar to the phonons that appear in the flux-line lattice of Abrikosov vortices in type-II superconductors: these are the Goldstone modes associated with the breaking of translational and rotational transverse symmetries by the lattice of vortices. We expect that these phonon modes hybridize with quantum fluctuations over the condensed $W$ field and reveal themselves in the $W$ boson spectrum.

\vskip 1mm
\begin{acknowledgments}
\paragraph*{\bf Acknowledgments.}
The work of MNC was funded by the EU's NextGenerationEU instrument through the National Recovery and Resilience Plan of Romania - Pillar III-C9-I8, managed by the Ministry of Research, Innovation and Digitization, within the project entitled ``Facets of Rotating Quark-Gluon Plasma'' (FORQ), contract no.~760079/23.05.2023 code CF 103/15.11.2022.
VAG was supported within the grant of the Ministry of Science and Higher Education of the Russian Federation (FEFU Program "PRIORITY 2030", topic No. ASP-25-03-1.03-0019).
AVM has been supported by Grant No. FZNS-2024-0002 of the Ministry of Science and Higher Education of Russia.
AVM thanks the Institut Dennis Poisson (Tours, France) for the kind hospitality and acknowledges the support of the Le Studium (Orl\'eans, France) research professorship.
The numerical simulations were performed at the computing cluster of Far Eastern Federal University, the equipment of Shared Resource Center "Far Eastern Computing Resource" IACP FEB RAS~\cite{IACPurl} and at computing resources of the Federal collective usage center Complex for Simulation and Data Processing for Mega-science Facilities at NRC "Kurchatov Institute"~\cite{NRCKIurl}.
\end{acknowledgments}

\bibliography{electroweak}

\end{document}